\documentclass[letterpaper,twocolumn,prl,aps,superscriptaddress,floatfix]{revtex4-2}
\usepackage{mathrsfs}
\usepackage{mathptmx}

\usepackage[latin9]{inputenc}
\usepackage{amsmath}
\usepackage{graphicx}
\usepackage{esint}
\usepackage{diagbox}
\usepackage{multirow}
\usepackage{algpseudocode}
\usepackage{algorithm}
\usepackage{ulem}

\makeatletter
\pdfpageheight\paperheight
\pdfpagewidth\paperwidth

\@ifundefined{textcolor}{}
{%
 \definecolor{BLACK}{gray}{0}
 \definecolor{WHITE}{gray}{1}
 \definecolor{RED}{rgb}{1,0,0}
 \definecolor{GREEN}{rgb}{0,1,0}
 \definecolor{BLUE}{rgb}{0,0,1}
 \definecolor{CYAN}{cmyk}{1,0,0,0}
 \definecolor{MAGENTA}{cmyk}{0,1,0,0}
 \definecolor{YELLOW}{cmyk}{0,0,1,0}
}
\usepackage{xcolor}
\usepackage{soul}

\usepackage{soul}

\newcommand{\ket}[1]{\ensuremath{\left|#1\right\rangle}}
\definecolor{blue}{rgb}{0,0,1}
\definecolor{red}{rgb}{1,0,0}
\definecolor{green}{rgb}{0,1,0}

\usepackage[unicode=true,pdfusetitle,
 bookmarks=true,bookmarksnumbered=false,bookmarksopen=false,
 breaklinks=false,pdfborder={0 0 0},pdfborderstyle={},backref=false,colorlinks=true]
 {hyperref}
\hypersetup{
 colorlinks,linkcolor=red,citecolor=blue}

\makeatother

\begin{document}
\title{All-optical control and multiplexed readout of multiple superconducting qubits}

\author{Xiaoxuan~Pan}
\thanks{These three authors contributed equally to this work.}
\affiliation{Center for Quantum Information, Institute for Interdisciplinary Information Sciences, Tsinghua University, Beijing 100084, China}

\author {Chuanlong~Ma}
\thanks{These three authors contributed equally to this work.}
\affiliation{Center for Quantum Information, Institute for Interdisciplinary Information Sciences, Tsinghua University, Beijing 100084, China}

\author {Jia-Qi~Wang}
\thanks{These three authors contributed equally to this work.}
\affiliation{Laboratory of Quantum Information, University of Science and Technology of China, Hefei, Anhui 230026, China}

\author {Zheng-Xu~Zhu}
\affiliation{Laboratory of Quantum Information, University of Science and Technology of China, Hefei, Anhui 230026, China}

\author{Linze~Li}
\affiliation{School of Information Science and Technology, ShanghaiTech University, Shanghai, China}

\author {Jiajun Chen}
\affiliation{Center for Quantum Information, Institute for Interdisciplinary Information Sciences, Tsinghua University, Beijing 100084, China}

\author {Yuan-Hao~Yang}
\affiliation{Laboratory of Quantum Information, University of Science and Technology of China, Hefei, Anhui 230026, China}

\author {Yilong~Zhou}
\affiliation{Center for Quantum Information, Institute for Interdisciplinary Information Sciences, Tsinghua University, Beijing 100084, China}

\author {Jia-Hua~Zou}
\affiliation{Laboratory of Quantum Information, University of Science and Technology of China, Hefei, Anhui 230026, China}

\author {Xin-Biao~Xu}
\thanks{E-mail: xbxuphys@ustc.edu.cn}
\affiliation{Laboratory of Quantum Information, University of Science and Technology of China, Hefei, Anhui 230026, China}

\author {Weiting~Wang}
\affiliation{Center for Quantum Information, Institute for Interdisciplinary Information Sciences, Tsinghua University, Beijing 100084, China}

\author{Baile~Chen}
\affiliation{School of Information Science and Technology, ShanghaiTech University, Shanghai, China}

\author{Haifeng~Yu}
\affiliation{Beijing Academy of Quantum Information Sciences, Beijing 100193, China}
\affiliation{Hefei National Laboratory, Hefei 230088, China}

\author{Chang-Ling~Zou}
\thanks{E-mail: clzou321@ustc.edu.cn}
\affiliation{Laboratory of Quantum Information, University of Science and Technology of China, Hefei, Anhui 230026, China}
\affiliation{Hefei National Laboratory, Hefei 230088, China}

\author{Luyan~Sun}
\thanks{E-mail: luyansun@tsinghua.edu.cn}
\affiliation{Center for Quantum Information, Institute for Interdisciplinary Information Sciences, Tsinghua University, Beijing 100084, China}
\affiliation{Hefei National Laboratory, Hefei 230088, China}


\begin{abstract}
\textbf{Superconducting quantum circuits operate at millikelvin temperatures, typically requiring independent microwave cables for each qubit for connecting room-temperature control and readout electronics. However, scaling to large-scale processors hosting hundreds of qubits faces a severe input/output (I/O) bottleneck, as the dense cable arrays impose prohibitive constraints on physical footprint, thermal load, wiring complexity, and cost. Here we demonstrate a complete optical I/O architecture for superconducting quantum circuits, in which all control and readout signals are transmitted exclusively via optical photons. Employing a broadband traveling-wave Brillouin microwave-to-optical transducer, we achieve simultaneous frequency-multiplexed optical readout of two qubits. Combined with fiber-integrated photodiode arrays for control signal delivery, this closed-loop optical I/O introduces no measurable degradation to qubit coherence times, with an optically driven single-qubit gate fidelity showing only a 0.19\% reduction relative to standard microwave operation. These results establish optical interconnects as a viable path toward large-scale superconducting quantum processors, and open the possibility of networking multiple superconducting quantum computers housed in separate dilution refrigerators through a centralized room-temperature control infrastructure.}

\end{abstract}

\maketitle
\vskip 0.5cm
\noindent \textbf{\large{}Introduction}{\large\par}
\noindent
The discovery of macroscopic quantum coherence in superconducting circuits has enabled the development of quantum bits based on Josephson junctions~\cite{devoret1985measurements,Clarke2008}, establishing superconducting processors as one of the leading platforms for universal quantum computation~\cite{Acharya2025Quantumerrorcorrection,Gao2025EstablishingNewBenchmark,2025Observationtopologicalprethermal,He2025ExperimentalQuantumError,nwaf246}. These circuits must operate at millikelvin temperatures, with each qubit requiring multiple coaxial cables that traverse successive thermal stages to room-temperature electronics for control, readout, and flux biasing~\cite{Krinner_2019,PRXQuantum040202}. Early studies of superconducting processors focused on manipulating a small number of qubits and validating fundamental quantum phenomena, with performance limited primarily by materials, fabrications, and circuit designs. Recent advances in these areas have extended the frontier to processors hosting hundreds of qubits~\cite{Acharya2025Quantumerrorcorrection,Gao2025EstablishingNewBenchmark,2025Observationtopologicalprethermal,He2025ExperimentalQuantumError}, but have simultaneously exposed a signal input/output (I/O) bottleneck that now threatens further scaling~\cite{Raicu2025,Croot2025}. Within a single dilution refrigerator, the dense array of coaxial cables imposes severe constraints on the number of individually addressable qubits~\cite{joshi2022scalingsuperconductingquantumcomputers}. For distributed architectures comprising multiple refrigerators, substantial microwave signal attenuation over distance impedes efficient routing, synchronization, and data exchange between processors.

Optical interconnects offer a compelling alternative as commercial optical fibers can potentially circumvent the limitations of microwave cables by providing negligible thermal conductance, compact footprints, and terahertz-scale bandwidth. For the control link, cryogenic photodiodes~\cite{Lecocq2021Controlreadoutsuperconducting,Multani2024Quantumlimitssuperconducting,Li2024Opticaltransmissionmicrowave,Xu2025Manipulationstransmonqubit} and coherent optical-to-microwave {transducers}~\cite{Warner2025Coherentcontrolsuperconducting} have been {used} to convert intensity-modulated light into  microwave pulses, enabling high-fidelity single-qubit gates. Recent demonstrations of parallel multi-qubit control and long-distance signal delivery have further validated the scalability of this approach~\cite{Ma2025}. For the readout link, coherent microwave-to-optical (M2O) transducers~\cite{Han2021Microwaveopticalquantum}  exploiting electro-optic, optomechanical, and piezo-optomechanical interactions have successfully upconverted qubit signals to the telecom band, validating an optical output channel for qubits~\cite{Mirhosseini2020Superconductingqubitoptical,Youssefi2021cryogenicelectrooptic,Delaney2022Superconductingqubitreadout,Arnold2025Allopticalsuperconducting,Thiel2025Opticalreadoutsuperconducting}. However, all prior coherent transducers have relied on discrete cavity modes, restricting their bandwidth to a narrow window compatible with only a single readout frequency. Consequently, demonstrations to date have been limited to single-qubit readout, and a complete all-optical I/O loop, where both control and readout signals propagate exclusively through optical fibers, has remained elusive.

Here, we demonstrate a complete optical I/O architecture for superconducting circuits. By employing a traveling-wave Brillouin M2O transducer~\cite{Wang2025} operating in a continuous-band regime, all-optical readout of multiple superconducting qubits using a single transducer is demonstrated. We achieve optical readout with a microwave bandwidth exceeding 200\,MHz, {substantially surpassing prior devices}, which allows further scale-up of the number of frequency-multiplexed qubits through a single device. Combining this broadband transducer with fiber-integrated uni-traveling-carrier (UTC) photodiode arrays~\cite{Li10475411,Song:21,RN22} for optical control delivery, we realize the first complete all-optical I/O architecture for superconducting quantum circuits, with all signals connecting room-temperature electronics and the refrigerator being transmitted via optical fibers. This closed-loop optical interface introduces no measurable degradation to qubit coherence times, and achieves $99.5\%$ single-qubit gate fidelity. By simultaneously advancing conversion bandwidth, multiplexing capability, and closed-loop operation, this work validates the optical links for addressing the I/O bottleneck confronting large-scale superconducting quantum computation.

\begin{figure}[t]
    \centering
    \includegraphics[scale=1]{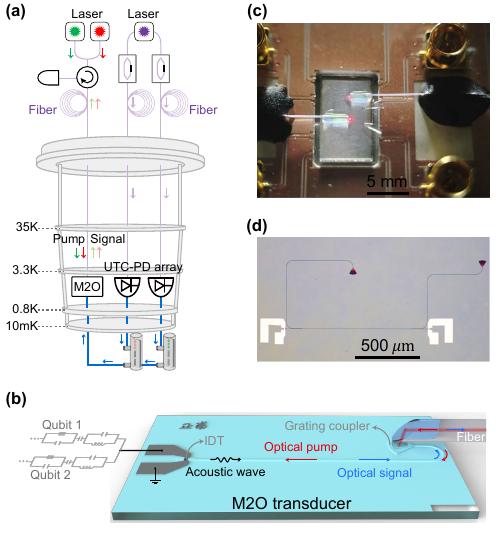}
    \caption[]{\textbf{The concept of cable-free optical I/O architecture for a superconducting quantum system.} \textbf{(a)} Schematic of the all-optical closed-loop link of qubit drive and readout. \textbf{(b)} Schematic of the traveling-wave Brillouin microwave-to-optical (M2O) transducer. \textbf{(c)} Photograph of the packaged M2O transducer. The microwave port is wire-bonded to a printed circuit board. The optical fiber is fixed onto the sample using fiber optic adhesive (black). The optical fiber end is located above the grating coupler (red). \textbf{(d)} Photograph of the structure of the M2O transducer.}
    \label{fig:Fig1}
\end{figure}

\smallskip{}

\noindent \textbf{\large{}Results}{\large\par}

\smallskip{}
\noindent\textbf{All-optical I/O architecture}

\noindent

Figure~\ref{fig:Fig1}(a) illustrates the optical I/O architecture for superconducting circuits. The key feature is the deployment of optical fibers from room temperature environment into the dilution refrigerator to replace traditional microwave transmission lines. Two complementary chips are fabricated and mounted on the 3.3\,K stage of a dilution refrigerator to interface optical signal and superconducting circuits: one chip integrates UTC photodiodes to convert optical signals into microwave pulses for control (downlink)~\cite{Ma2025}, while the other chip incorporates M2O transducers for qubit readout (uplink) [Figs.~\ref{fig:Fig1}(b)-(d)]. In our implementation, optical fibers of $\sim 30~\mathrm{m}$ in length are employed to connect the laser sources to each chip, while short microwave cables $\sim1~\mathrm{m}$ interconnect these chips to the qubits at millikelvin stage.

Realizing the closed-loop optical interface with potential scalability requires both the downlink and uplink to operate with sufficient bandwidth to address multiple qubits through shared optical fibers. For the control downlink, this requirement is fullfilled with UTC photodiodes having an intrinsically broad electrical bandwidth extending 10\,GHz, which readily accommodates multiple frequency-multiplexed control tones~\cite{Ma2025}. This allows precise qubit control gates and readout probe tones to be synthesized locally at the cryogenic stage through a single optical fiber. However, the multiplexing for readout uplink presents a critical challenge for conventional M2O transducers, which rely on discrete cavity modes in both the microwave and optical domains and only support a single conversion channel with a narrow bandwidth of a few megahertz.

To overcome this limitation, we implement a traveling-wave Brillouin M2O transducer {within a compact footprint}~\cite{Yang2025}. As illustrated in Fig.~\ref{fig:Fig1}(b), the transduction process proceeds through two cascaded interactions. First, an interdigital transducer (IDT) converts incoming microwave photons into traveling acoustic phonons via the piezoelectric effect in a lithium niobate ridge waveguide on a sapphire substrate. Subsequently, under an optical pump, these phonons are coherently converted into optical photons as an optical sideband. The co-propagation of acoustic and optical waves in the straight waveguide leads to phase-matched interactions over a continuous operation bandwidth for the input microwave signal~\cite{Wang2025,Yang2025}. Figures~\ref{fig:Fig1}(c) and \ref{fig:Fig1}(d) show photographs of the packaged M2O transducer, where the microwave port is wire-bonded to a printed circuit board and the optical fiber is aligned to a grating coupler using fiber-optic adhesive. 

\begin{figure}[t]
	\includegraphics{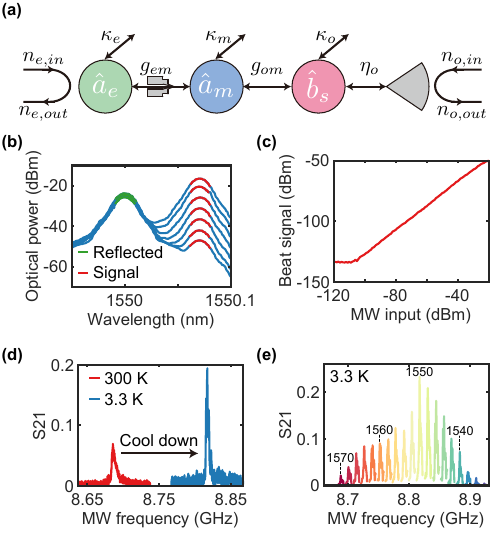}
	\caption{\textbf{Characterization of the broadband traveling-wave M2O transducer.} \textbf{(a)} Schematic of the conversion process between the microwave mode ($\hat{a}_\text{e}$) and the optical mode ($\hat{b}_\text{s}$). $\kappa_{e},\kappa_m$, and $\kappa_o$ represent the coupling rates of each mode to the external environment. $n_{in(out)}$ denotes the photon number of the input (output) field. \textbf{(b)} Reflected optical spectrum from the transducer at room temperature for different input microwave powers (from $-15\,\text{dBm}$ to $+15\,\text{dBm}$ in $5\,\text{dBm}$ increments), showing both reflected pump ($1550.00\,\text{nm}$, green line) and the generated Stokes signal ($1500.07\,\text{nm}$, red line). \textbf{(c)} Heterodyne beat note power as a function of input microwave power, with a fixed pump ($100\,\text{mW}, 1550\,\text{nm}$). The reflected light is amplified to $5\,\text{dBm}$ using an Erbium-Doped Fiber Amplifier (EDFA) before detection by a high-speed photodetector (HPD) to obtain the beat note signal. \textbf{(d)} Microwave spectrum of the beat note signal for a fixed pump ($100\,\text{mW}, 1550\,\text{nm}$) before and after cooling in the dilution refrigerator. The M2O transducer is mounted on the $3.3\,\text{K}$ stage. \textbf{(e)} Microwave spectrum of the beat note signal at $3.3\,\text{K}$ for different pump wavelengths ($100\,\text{mW}$) swept from $1534\,\text{nm}$ to $1570\,\text{nm}$ in $2\,\text{nm}$ increments.}
	\label{fig:fig2}
\end{figure}

\smallskip{}
\noindent\textbf{Broadband traveling-wave M2O transducer}

\noindent
Figure~\ref{fig:fig2}(a) illustrates the working principle of our traveling-wave Brillouin M2O transducer through the signal conversion flow among the electric ($\hat{a}_\text{e}$), mechanical ($\hat{a}_\text{m}$), and optical ($\hat{b}_\text{s}$) modes. The piezoelectric interaction via the IDT is described by a beam-splitter-like Hamiltonian $\hat{H}_{\text{em}} = \hbar g_{\text{em}} (\hat{a}_\text{e}^\dagger \hat{a}_\text{m} + \hat{a}_\text{e} \hat{a}_\text{m}^\dagger)$, with $g_{\text{em}}$ being the piezoelectric coupling rate. Subsequently, under an external optical pump field $\hat{b}_p$, the Brillouin scattering process is governed by the optomechanical Hamiltonian $\hat{H}_{\text{om}} = \hbar g_{\text{om}} (\hat{b}_\text{p}^\dagger \hat{b}_\text{s} \hat{a}_\text{m} + \hat{b}_\text{p} \hat{b}_\text{s}^\dagger \hat{a}_\text{m}^\dagger)$, with $g_{\text{om}}$ denoting the single-photon optomechanical coupling strength. By fulfilling the phase-matching condition for the Stokes/anti-Stokes processes (energy conservation $\omega_{\text{pump}} = \omega_{\text{signal}} \pm \omega_{\text{phonon}}$), the Brillouin interaction~\cite{Brillouin2019} enables coherent linear/bilinear conversion between the acoustic ($\hat{a}_\text{m}$) and optical ($\hat{b}_\text{s}$) domains. It is worth noting that our device supports both Stokes and anti-Stokes processes simply by reversing the pump incidence direction through different grating coupler ports, as indicated in Fig.~\ref{fig:Fig1}(d). In this work, we operate in the Stokes regime, which exhibits higher optical coupling efficiency at the corresponding input port.

We first characterize the transducer at room temperature by sending microwave signals to the IDT with varying powers under a fixed optical pump (100\,mW at 1550\,nm). Figure~\ref{fig:fig2}(b) shows the optical spectra of collected light, consisting of a constant pump reflection alongside a red-shifted Stokes sideband whose peak power (highlighted by the red curves) scales proportionally with the input microwave power. Since the measurement through an optical spectrum analyzer is limited by the background of reflected optical pump, we also employ a heterodyne detection scheme to resolve the weak optical signals. The reflected pump and scattered signal are amplified via an erbium-doped fiber amplifier (EDFA) and directed to a high-speed photodiode to generate a beat note. This beat note, recorded by a microwave spectrum analyzer, provides stable and reproducible measurement of the converted optical signal power. Figure~\ref{fig:fig2}(c) shows the extracted beat note power as a function of input microwave power, confirming the linear transduction process. We derive an M2O transduction efficiency of $3.2\times 10^{-7}/\mathrm{W}$, defined as the ratio of output optical to input microwave photon flux per pump power.

\begin{figure*}[ht]
    \centering
    \includegraphics[scale=1]{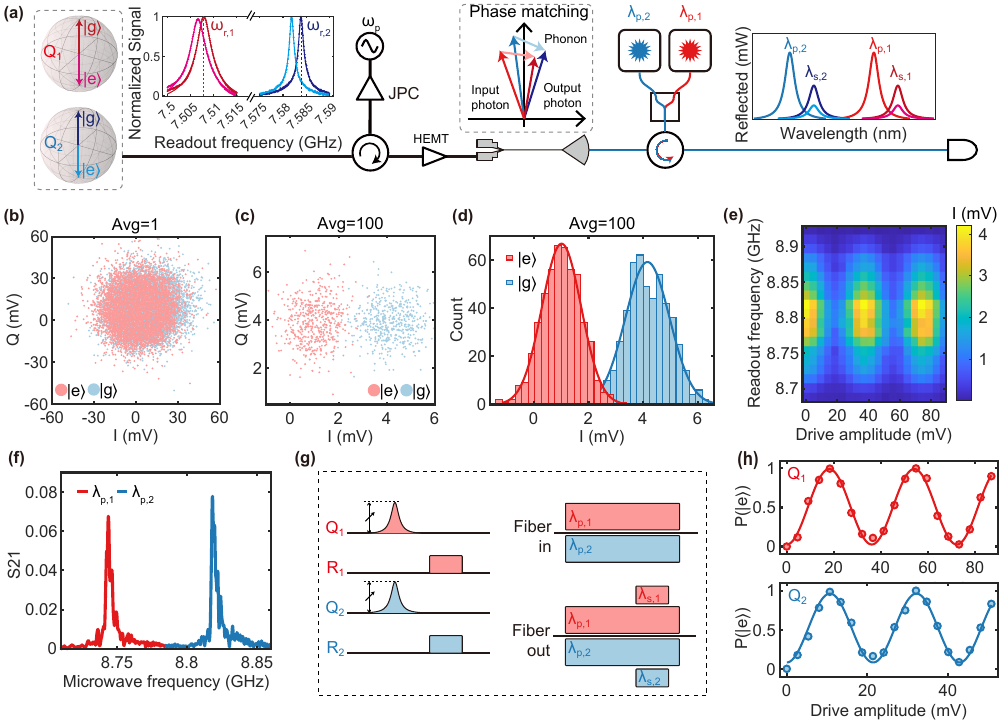}
    \caption{\textbf{Simultaneous readout of multiple qubits via the M2O transducer.}\textbf{(a)} Diagram of the simultaneous readout. The readout signals of two transmon qubits ($\omega_{r,1},\omega_{r,2} =7.509\,\mathrm{GHz}, 7.584\,\mathrm{GHz}$) are upconverted to $8.743\,\mathrm{GHz},8.818\,\mathrm{GHz}$ by a Josephson parametric converter (JPC) using a $1.234\,\mathrm{GHz}$ microwave pump. These microwave readout signals are then simultaneously converted into different optical signals by the transducer with two pump lights ($\lambda_{p,1}=1550.00\,\mathrm{nm},\lambda_{p,2}=1560.95\,\mathrm{nm}$). \textbf{(b) and (c)} IQ scatter diagrams of the single-qubit optical readout of qubit $\mathrm{Q}_1$, averaged over 1 and 100 measurements, respectively. The readout pulse length is $10\,\mu \mathrm{s}$. \textbf{(d)} Histogram of the I quadrature averaged over 100 measurements. \textbf{(e)} Optical readout of a power Rabi experiment on transmon qubit $\mathrm{Q}_1$. The optical pump wavelength and the corresponding microwave frequency are tuned to enable an optical readout over a bandwidth of $200\,\mathrm{MHz}$. The color map represents the amplitude of the beat note signal.  \textbf{(f)} Microwave spectrum of the beat note signal when the transducer is pumped simultaneously by two pumps ($\lambda_1=1550.00\,\mathrm{nm},\lambda_2=1560.95\,\mathrm{nm}$). \textbf{(g)} Optical and microwave pulse sequences for the simultaneous optical readout. \textbf{(h)} Simultaneous optical readout of the power Rabi experiment for two qubits, with each data point averaged over $5\times 10^4$ measurements.}
    \label{fig:Fig3}
\end{figure*}

The frequency response of the device is further characterized using a vector network analyzer (VNA). Figure~\ref{fig:fig2}(d) compares the microwave-optical scattering coefficients (S21) measured at room temperature (300\,K) and cryogenic temperature (3.3\,K) with a fixed optical pump. At room temperature, the S21 spectrum reveals a conversion channel with a 3\,dB-bandwidth of $2.63 \,\mathrm{MHz}$. When the system is cooled, the dispersion and attenuation of the phononic waveguide change, shifting the central frequency of the conversion channel by $131.56\,\mathrm{MHz}$ and narrowing its bandwidth to $0.88\,\mathrm{MHz}$, while the amplitude of optical signal is enhanced by a factor of $2.8$. Consequently, the efficiency of the transducer increases to $2.5\times 10^{-6}/\mathrm{W}$ in the cryogenic environment.

Crucially, our traveling-wave design enables broadband tunability of the conversion channel. As shown in Fig.~\ref{fig:fig2}(e), by sweeping the optical pump wavelength across the telecom C-band, the central frequency of the conversion channel shifts accordingly, covering a total microwave bandwidth exceeding 200\,MHz. This frequency range is currently limited by the bandwidth of the IDT and grating couplers rather than by fundamental constraints of the Brillouin interaction, indicating that even larger bandwidths are achievable. This cavity-free architecture permits parallel operation of multiple conversion channels at arbitrarily selected frequencies by sending multiple pump wavelengths simultaneously, enabling the multiplexed qubit readout demonstrated in the following. We note that the transduction efficiency can be further boosted by incorporating traveling-wave cavities~\cite{Yang2024ProposalBrillouinmicrowave}, and preliminary investigations have confirmed that multiple parallel channels at fixed frequencies remain accessible~\cite{Yang2025b}.

\begin{figure*}
    \centering
    \includegraphics[scale=1]{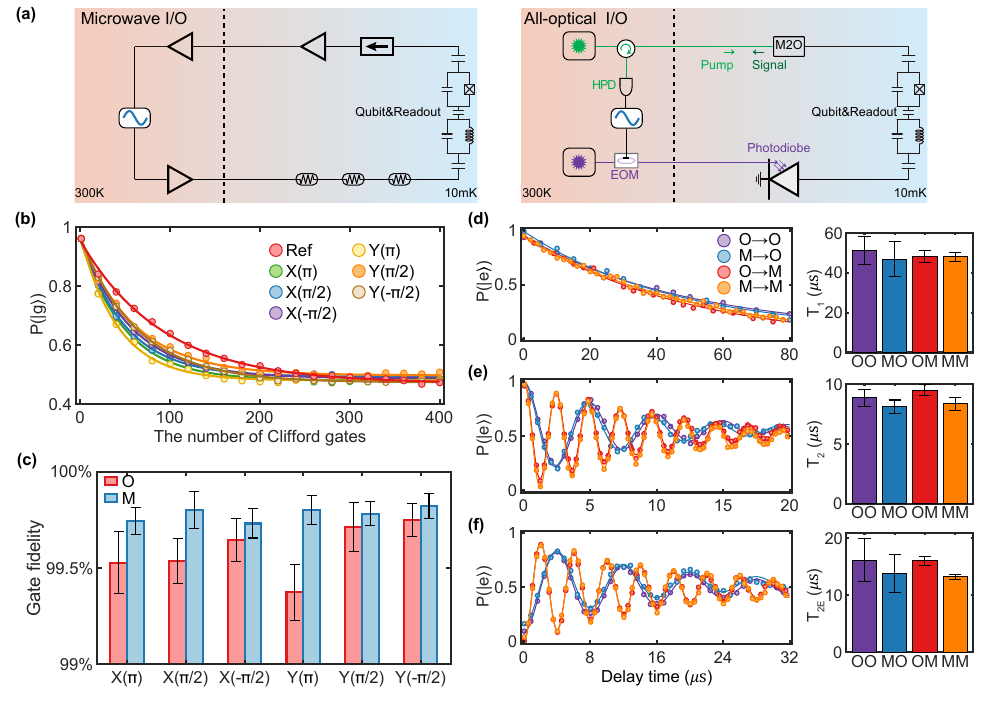}
    \caption[]{\textbf{All-optical control and readout of superconducting qubits.} \textbf{(a)} Comparison between standard microwave I/O and all-optical I/O. \textbf{(b)} Random Benchmarking experiment under optical drive. Both the qubit control and readout pulses are generated via the optical downlink. \textbf{(c)} Single-qubit gate fidelity under optical and microwave drive. All gates are implemented using $120$\,ns Gaussian pulses. \textbf{(d), (e), (f)} Measurements of $T_1$, $T_2$, and $T_\mathrm{2E}$ of the superconducting qubit ($\mathrm{Q}_2$) under different combinations of drive and readout methods: optical drive with optical readout (purple), microwave drive with optical readout (blue), optical drive with microwave readout (red), and microwave drive with microwave readout (orange). $P(\ket{g})$ and $P(\ket{e})$ are the measured populations of the qubit in the ground and excited states, respectively. In the Ramsey experiment, the two $\pi/2$ pulses are applied with a detuning of $0.20$ \,MHz relative to the qubit frequency for optical readout and $0.42$\,MHz for microwave readout.}
    \label{fig:Fig4}
\end{figure*}

\smallskip{}
\noindent\textbf{Parallel readout of multiple qubits}

\noindent
Figure~\ref{fig:Fig3}(a) depicts the experimental setup for the multiplexed optical readout of two superconducting qubits. In the dispersive readout regime, each qubit induces a state-dependent shift to the readout cavity, which encodes the qubit state information ($\ket{g}$ or $\ket{e}$) in the amplitude and phase of a reflected microwave probe pulse~\cite{Blais2021Circuitquantumelectrodynamics}. Since the transducer operates in an inherent coherent manner, the quadrature information encoded in the microwave domain is faithfully transferred to the optical domain, and eventually be recovered through heterodyne detection. To bridge the frequency gap between the qubit readout frequencies ($\sim 7.5$\,GHz) and the optimal operation window of the M2O transducer ($\sim 8.8$\,GHz), we employ a flux-pumped Josephson parametric converter (JPC) to upconvert the readout tones~\cite{Eichler2014QuantumLimitedAmplification,Grebel2021Fluxpumpedimpedance}. We note that this frequency bridging is specific to the current device parameters and can be eliminated in future designs by matching the transducer bandwidth directly to the readout cavity frequencies. {Additionally, a high-electron-mobility-transistor (HEMT) amplifier at 3.3~K provides gain to enhance the transduction signal-to-noise ratio.} The upconverted {and amplified} microwave photons are then directed to the M2O transducer, where they are coherently mapped to optical sidebands via the Stokes process.

We first validate the optical readout link using a single qubit ($\mathrm{Q}_1$). Figures~\ref{fig:Fig3}(b) and \ref{fig:Fig3}(c) display the quadrature distributions of the optical beat note signal for the qubit prepared in the ground $|g\rangle$ and excited $|e\rangle$ states. For {individual} acquisition, the distributions for the two states overlap substantially due to the limited transduction efficiency. However, after averaging over 100 repetitions, the distributions become well separated, confirming that the phase information of the microwave photon is faithfully preserved when transferred through the optical link. The corresponding histogram projected onto the I quadrature [Fig.~\ref{fig:Fig3}(d)] shows well-resolved peaks for the two qubit states. To further characterize the readout performance, we measure Rabi oscillations by varying the amplitude of a resonant drive pulse applied to $\mathrm{Q}_1$ prior to optical readout. Figure~\ref{fig:Fig3}(e) presents these oscillations as a function of both drive amplitude and upconverted readout frequency. The observed signal-to-noise ratio is consistent with the measured system transduction efficiency and confirms the viability of optical readout for qubit state discrimination {over a $200$\,MHz microwave bandwidth}.

The hallmark of our optical I/O architecture is the ability to read out multiple qubits using a single broadband traveling-wave M2O transducer. As illustrated in Fig.~\ref{fig:Fig3}(a), we implement frequency-division multiplexing by driving the transducer with two distinct optical pump tones at wavelengths $\lambda_{p,1}=1550.00\,\mathrm{nm}$ and $\lambda_{p,2}=1560.95\,\mathrm{nm}$). Each pump selectively addresses a specific microwave readout frequency determined by the phase-matching condition~\cite{Wang2025}, establishing two independent conversion channels at $8.743\,\mathrm{GHz}$ and $8.818\,\mathrm{GHz}$, respectively. As shown in Fig.~\ref{fig:Fig3}(f), by simultaneously pumping at $\lambda_{p,1}$ and $\lambda_{p,2}$, we establish dual-channel M2O transduction via a single device for the simultaneous readout of both qubits. Finally, we perform simultaneous time-domain measurements. The experimental pulse sequence for simultaneous Rabi measurements, shown in Fig.~\ref{fig:Fig3}(g), illustrates that both qubits ($\mathrm{Q}_1$ and $\mathrm{Q}_2$) are independently driven, followed by simultaneous readout of their corresponding cavities ($\mathrm{R}_1$ and $\mathrm{R}_2$). Figure~\ref{fig:Fig3}(h) shows the resulting synchronized Rabi oscillations from parallel optical readout of two driven qubits via a single optical link, confirming the parallel processing capability of our I/O architecture.

\smallskip{}
\noindent\textbf{Closed-loop optical I/O}

\noindent

To complete the closed-loop optical I/O architecture, we combine the optical readout uplink with an optical downlink for qubit control. Figure~\ref{fig:Fig4}(a) contrasts the standard coaxial I/O with our all-optical architecture. By replacing the bulky coaxial attenuation chains with optical fibers, we eliminate the primary thermal and spatial overheads associated with conventional microwave wiring. In our scheme, control signals are applied onto an optical carrier via a room-temperature electro-optic modulator (EOM) and transmitted to the UTC photodiode array. The EOM is biased at its quadrature point to ensure linear electrical-to-optical conversion. Leveraging the wide bandwidth of the UTC photodiodes, this single optical channel synthesizes both the $4.954$\,GHz pulses for qubit gates and the $7.584$\,GHz probe tones for readout resonators directly at the millikelvin stage. We validate the control fidelity using single-qubit randomized benchmarking (RB)~\cite{Magesan2012EfficientMeasurementQuantum}. As shown in Figs.~\ref{fig:Fig4}(b) and \ref{fig:Fig4}(c), the optical drive achieves an average gate fidelity of 99.59\%, exhibiting only a minor reduction of $0.19\%$ compared to the coaxial baseline. This result confirms that the optical-to-microwave conversion at the UTC photodiode introduces negligible noise, making it a viable alternative to metallic transmission lines for high-fidelity qubit manipulation.

The ultimate test for an optical I/O interface is whether the simultaneous operation of optical drive (downlink) and readout (uplink) introduce decoherence. We integrate both subsystems to form a fully optical closed loop, where all input and output signals connecting the room-temperature electronics and the millikelvin qubits are mediated exclusively by optical photons through fibers. We systematically characterize the qubit coherence properties (longitudinal relaxation time $T_1$, Ramsey dephasing time $T_2$, and Hahn-echo coherence time $T_\mathrm{2E}$) under four distinct I/O configurations: purely microwave (M-M), microwave drive with optical readout (M-O), optical drive with microwave readout (O-M), and fully optical (O-O). The results, summarized in Figs.~\ref{fig:Fig4}(d)-(f), reveal a striking consistency: the qubit coherence times remain virtually unchanged across all configurations. Specifically, even in the fully optical (O-O) regime, where the refrigerator is subjected to continuous optical illumination for both readout upconversion and control drive, we observe no measurable degradation in $T_1$ ($\sim 51.0\,\mu \text{s}$), $T_2$ ($\sim 8.8\,\mu \text{s}$) or $T_\mathrm{2E}$ ($\sim 16.2\,\mu \text{s}$). This result demonstrates that our fiber-packaged transducers and UTC photodiodes are well-thermalized, validating the feasibility of this all-optical architecture for high-coherence superconducting quantum processors.

\smallskip{}

\noindent \textbf{\large{}Discussion}{\large\par}

\noindent
A compact closed-loop optical I/O system for controlling and reading out superconducting qubits has been implemented, where all microwave cables to deliver signals are replaced by optical fibers, providing a promising solution to the critical wiring bottleneck in large-scale superconducting quantum processors. The demonstrated broadband traveling-wave microwave-to-optical (M2O) transducer offers a bandwidth over 200\,MHz, supporting more than 40 independent channels with 5\,MHz spacing through a single device. The main challenges to further scaling up the number of parallel operating qubits are twofold: the limited cooling power of the refrigerator for handling the relatively high optical pump power and the complexity of multiplexing optical wavelengths for each individual channel. Both challenges can be substantially mitigated through the advancements in photonic chips. First, the efficiency of the Brillouin M2O transducer can be significantly improved by introducing optical and acoustic traveling-wave cavities, with a projected conversion efficiency reaching $10^{-2}/\mathrm{W}$~\cite{Yang2024ProposalBrillouinmicrowave,Yang2025b}. This would significantly reduce the pump power and enhance the readout signal-to-noise ratio. Second, multi-wavelength optical pump sources, along with multiplexing and modulation of laser tones, can be implemented using on-chip photonic components~\cite{Sun2023,Alexander2025}, such as microcomb sources, arrayed waveguide grating, and integrated acousto-optic or electro-optic modulators. With a cooling power of $1\,\mathrm{W}$ at the 4\,K stage, it is possible to scale up the number of parallel operating qubits to 100 using a single M2O transducer and a single optical fiber.

The coherent transduction demonstrated in this work not only addresses classical signal routing but also establishes the infrastructure for quantum communication between superconducting processors. When the transduction efficiency reaches $10^{-2}$ with sufficiently low added noise, the same M2O device can mediate entanglement generation between remote qubits or superconducting modes~\cite{Zhong2020,Krastanov2021,Zhao2025Quantumenabledmicrowave,Zou2025}. While this work focuses on superconducting qubits, the optical I/O architecture addresses a universal challenge shared by all cryogenic quantum systems, such as semiconductor quantum dots~\cite{Chatterjee2021}. It thus offers a scalable pathway to overcome the I/O bottleneck and enables distributed quantum information processing.

\smallskip{}

\noindent \textbf{\large{}{}Acknowledgment}{\large\par}

\noindent This work was funded by the Quantum Science and Technology-National Science and Technology Major Project (Grant Nos.~2024ZD0301500 and 2021ZD0300200), the National Natural Science Foundation of China (Grant Nos.~92265210, 123B2068, 92165209, 92365301, 12474498, 92565301, 12374361, 12404567, and 12293053). We also acknowledge the support from the Fundamental Research Funds for the Central Universities and USTC Research Funds of the Double First-Class Initiative. The UTC photodiode chips were fabricated with support from the ShanghaiTech University Material and Device Lab (SMDL). The numerical calculations in this paper were performed on the supercomputing system in the Supercomputing Center of USTC, and this work was partially carried out at the USTC Center for Micro and Nanoscale Research and Fabrication.

\end{document}